\documentclass[twocolumn]{aastex631}
\usepackage{threeparttable}
\usepackage{amsmath}

\usepackage{xcolor}
\usepackage{tikz}
\usetikzlibrary{arrows,shapes,chains}

\newcommand{\grizy}{$g$, $r$, $i$, $z$ and $y$}
\newcommand{\bprp}{$G_{\rm BP}-G_{\rm RP}$}
\newcommand{\bpg}{$G_{\rm BP}-g$}
\newcommand{\rpr}{$G_{\rm RP}-r$}
\newcommand{\rpi}{$G_{\rm RP}-i$}
\newcommand{\rpz}{$G_{\rm RP}-z$}
\newcommand{\rpy}{$G_{\rm RP}-y$}
\newcommand{\feh}{$\rm [Fe/H]$}
\newcommand{\teff}{$T_{\rm eff}$}
\newcommand{\ebprp}{$E(G_{\rm BP}-G_{\rm RP})$}
\shorttitle{Validation and Improvement of the Pan-STARRS Photometric Calibration}
\shortauthors{Xiao et al.}
\graphicspath{{./}{figures/}}

\begin{document}
\title{Validation and Improvement of the Pan-STARRS Photometric Calibration with the Stellar Color Regression Method}
% \correspondingauthor{Haibo Yuan}
% \email{yuanhb@bnu.edu.cn}
% \author[0000-0001-8424-1079]{Kai Xiao}
% \affiliation{Department of Astronomy, Beijing Normal University, Beijing, 100875, People's Republic of China}
% \author[0000-0003-2471-2363]{Haibo Yuan}
% \affiliation{Department of Astronomy, Beijing Normal University, Beijing, 100875, People's Republic of China}
\author
{
Kai Xiao\altaffilmark{1}
Haibo Yuan\altaffilmark{1}
}
\altaffiltext{1}{Department of Astronomy, Beijing Normal University, Beijing, 100875, People's Republic of China; email: yuanhb@bnu.edu.cn}

\journalinfo{submitted to AJ}
\submitted{Received: 2021 December 28; Revised: 2022 February 08; Accepted: 2022 February 09}

\begin{abstract}
As one of the best ground-based photometric dataset, 
Pan-STARRS1 (PS1) has been widely used as the reference to calibrate other surveys.
In this work, we present an independent validation and re-calibration of the PS1 photometry using spectroscopic data from the LAMOST DR7 and photometric data from the corrected Gaia EDR3 with the Stellar Color Regression (SCR) method. 
Using per band typically a total of 1.5 million LAMOST-PS1-Gaia stars as standards, 
we show that the PS1 photometric calibration precisions in the $grizy$ filters 
are around $4\sim 5$ mmag when averaged over $20'$ regions.
However, significant large- and small-scale spatial variation of magnitude offset, 
up to over 1 per cent, probably caused by the calibration errors in the PS1, are found for all the $grizy$ filters.
The calibration errors in different filters are un-correlated, and are slightly larger for the $g$ and $y$ filters. We also detect moderate magnitude-dependent errors (0.005, 0.005, 0.005, 0.004, 0.003 mag per magnitude in the 14 -- 17 magnitude range for the $grizy$ filters, respectively) 
in the PS1 photometry by comparing with the 
Gaia EDR3 and other catalogs. The errors are likely caused by the systematic uncertainties in the PSF magnitudes. 
We provide two-dimensional maps to correct for such magnitude offsets in the LAMOST footprint at different spatial resolutions from $20'$ to $160'$. 
The results demonstrate the power of the SCR method in improving the calibration precision 
of wide-field surveys when combined with the LAMOST spectroscopy and Gaia photometry. 

\end{abstract}


% ==============================================================================
\section{Introduction} \label{sec:intro}

The current and next-generation wide-field imaging surveys such as the Sloan Digital Sky Survey (SDSS; \citealt{2000AJ....120.1579Y}), the Panoramic Survey Telescope and Rapid Response System (Pan-STARRS; \citealt{2002SPIE.4836..154K}), the Dark Energy Survey (DES; \citealt{2015AJ....150..150F,2016MNRAS.460.1270D,2018ApJS..239...18A}), the Javalambre Physics of the Accelerating Universe Astrophysical Survey (J-PAS; \citealt{2014arXiv1403.5237B}), the Wide Field Survey Telescope (WFST; \citealt{2016SPIE10154E..2AL}), the Chinese Space Station Telescope (CSST; \citealt{2018cosp...42E3821Z}), the Legacy Survey of Space and Time (LSST; \citealt{2019ApJ...873..111I}) and the Multi-channel Photometric Survey Telescope (Mephisto; Er et al. 2021, in preparation), 
are vital in modern astronomy in discovering and characterizing new objects and phenomena. 
While uniform and accurate photometric calibration play a central role in the wide-field surveys. 

Recently, some new approaches have been developed for the high-precision calibration of wide field surveys. 
The methods can be divided into categories of either: ``hardware-driven" or ``software-driven" (\citealt{Huang}). Approaches of the former category are based on better understanding of the wide-field imaging observations, and include such as the Ubercalibration method (\citealt{2008ApJ...674.1217P}), the Forward Global Calibration Method (FGCM; \citealt{2018AJ....155...41B}), and the Hypercalibration method (\citealt{2016ApJ...822...66F}); 
while those of the latter category are based on better understanding of stellar colors, such as the Stellar Locus Regression method (SLR; \citealt{2009AJ....138..110H}), the Stellar Locus method (SL; \citealt{2019A&A...631A.119L}), and the Stellar Color Regression method (SCR; \citealt{2015ApJ...799..133Y}). 

Owing to the rapid development of multi-fiber spectroscopic surveys, e.g., LAMOST (\citealt{2012RAA....12..735D}; \citealt{2014IAUS..298..310L}), we have entered into the era of millions of stellar spectra. In addition, with the modern template-matching and data-driven based stellar parameter pipelines (e.g., \citealt{2008AJ....136.2022L,2008AJ....136.2050L,2011RAA....11..924W}; \citealt{2015MNRAS.448..822X,2017MNRAS.464.3657X}), stellar atmospheric parameters, such as \teff, $\log g$, \feh, can be determined to a very high internal precision (e.g., \citealt{2021ApJ...909...48N}). As a result, stellar colors can now be accurately predicted based on the large-scale spectroscopic surveys. Using millions of spectroscopically observed stars as color standards, \citet{2015ApJ...799..133Y} first proposed the spectroscopy-based SCR method and performed precise color (re-)calibrations for the SDSS Stripe 82. Compared to the other ``software-driven'' methods, the SCR method fully accounts the effects of metallicity, surface gravity, and dust reddening on stellar colors.
When applied to the SDSS Stripe\,82 (\citealt{2007AJ....134..973I}), it achieved a precision of 2 -- 5 mmag in the SDSS colors. The method has also been applied to the Gaia Data Release 2 and Early Data Release 3 (EDR3) to correct for the magnitude/color-dependent systematic errors in 
the Gaia colors (\citealt{2021ApJ...909...48N,2021ApJ...908L..14N}), achieving an unprecedented precision of 1 mmag. 
Together with the high-precision photometry from Gaia,
the SCR method can further be used to predict stellar magnitudes accurately and perform 
photometric calibration. For example, \citet{2021ApJ...907...68H} have applied the 
method to recalibrate the DR2 of the SkyMapper Southern Survey (SMSS; \citealt{2018PASA...35...10W}), and 
find large zero-point offsets in the $uv$ bands.
\citet{Huang} have applied the method to the SDSS Stripe\,82 standard stars 
catalogs (\citealt{2007AJ....134..973I,2021MNRAS.505.5941T}), 
achieving a precision of 5 mmag in the $u$ band, and 2 mmag in the $griz$ bands.
Possible implementations of the SCR method under different situations and improvements 
are also discussed by \citet{Huang}.

As the first part of the Pan-STARRS (\citealt{2002SPIE.4836..154K, 2010SPIE.7733E..0EK}), 
Pan-STARRS1 (PS1; \citealt{2012ApJ...750...99T}) has imaged three quarters of the sky repeatly 
in five broadband filters ($g$, $r$, $i$, $z$, $y$).
Taking advantage of the large amount of over-lapping observations, 
PS1 photometry has been calibrated using the ubercalibration method to a precision 
better than 1 per cent \citep{2012ApJ...756..158S,2020ApJS..251....6M}. 
As one of the best ground-based photometric dataset, 
PS1 has been widely used as reference to calibrate other surveys,
including the SDSS survey (\citealt{2016ApJ...822...66F}), the Beijing–Arizona Sky Survey (BASS; \citealt{2017AJ....153..276Z,2018PASP..130h5001Z}), the J-PLUS (\citealt{2019A&A...631A.119L, 2021A&A...654A..61L}). 
It has also been used as cross-calibration of multiple photometric systems to improve the cosmological measurements with Type Ia supernovae (see \citealt{2015ApJ...815..117S,2021arXiv211203864B}). 

In this work, using the SCR method with the corrected photometric data from the Gaia EDR3 (\citealt{2021ApJ...908L..24Y}) and the spectroscopic date from LAMOST DR7, validation and improvement of the PS1 photometric calibration are performed. The paper is organized as follows. In Sections~\ref{sec:data} and \ref{sec:method}, we introduce the data used and the validation process with the SCR method in this work. The results are presented in Section~\ref{sec:result} and discussed in Section~\ref{sec:discussion}. Conclusions are given in Section~\ref{sec:conclusion}. 
% ==============================================================================
\section{Data} \label{sec:data}
\subsection{Pan-STARRS 1 Data Release 1} \label{sec:ps1}
The PS1 survey has imaged three quarters of the sky in five broadband filters ($g$, $r$, $i$, $z$, $y$), using its 1.8 meter telescope of a $3.3^{\circ}$ field of view (\citealt{2004AN....325..636H}) and 1.4 Gigapixel camera. Its first public data release (DR1) on 16 December 2016 contains the results of the third full reduction of the Pan-STARRS $3\pi$ Survey (\citealt{2020ApJS..251....6M}). The typical $5\sigma$ limiting magnitudes for 
point sources are (23.3, 23.2, 23.1, 22.3, 21.4) in the ($g$, $r$, $i$, $z$, $y$) bands, respectively (\citealt{2016arXiv161205560C}). PSF magnitudes, Kron magnitudes, and aperture magnitudes 
are provided in the PS1 DR1.  PSF magnitudes are obtained from fitting a predefined 
PSF model using maximum-likelihood methods, and mainly for stars. Kron magnitudes are mainly 
for extended sources. Aperture magnitudes measure the total flux for a point source based on integration 
over an aperture plus an extrapolation according to the PSF. In this work, PSF magnitudes are used as default.

\subsection{Gaia Early Data Release 3} \label{sec:gaia}
The EDR3 (\citealt{2021A&A...649A...1G,2021A&A...650C...3G}) of the European Space Agency (ESA)’s space mission Gaia (\citealt{2016A&A...595A...1G}) has delivered not only the best astrometric information but also the best photometric data for about 1.8 billion stars in $G$, $G_{\rm BP}$ and $G_{\rm RP}$ bands, in terms of full sky coverage, uniform calibration at mmag level, and small photometric errors for a very wide range of magnitudes. The overall calibration errors are less than 1 mmag for overall trend except for very blue and bright sources (\citealt{2021A&A...649A...3R}). More recently, \citet{2021ApJ...908L..24Y} carried out an independent validation of Gaia EDR3 photometry against about 10,000 Landolt standard stars using a machine-learning technique. They obtained magnitude-dependent corrections up to 10 mmag for the three Gaia bands. 
Hence, Gaia EDR3 magnitudes hereafter refer to those corrected by \citet{2021ApJ...908L..24Y}.

\subsection{LAMOST Data Release 7} \label{sec:lm}
The Large Sky Area Multi-Object Fiber Spectroscopic Telescope (LAMOST; \citealt{2012RAA....12.1197C}; \citealt{2012RAA....12..723Z}; \citealt{2012RAA....12..735D}; \citealt{2014IAUS..298..310L}) is a quasi-meridian reflecting Schmidt telescope with 4000 fibers, and has a field of view of 20 deg$^2$. Its Data Release 7 (hereafter DR7; see \citealt{2015RAA....15.1095L}) includes a total number of 10,640,255 low resolution spectra covering the whole optical wavelength range of 369 -- 910 nm at a spectral resolution of about 1800. The LAMOST Stellar Parameter Pipeline (LASP; \citealt{2011RAA....11..924W}) has been used to determine the basic stellar parameters including effective temperature \teff, surface gravity $\log g$ and metallicity \feh. The typical precision is 
about 110 K for \teff, 0.2 dex for $\log g$, and 0.1 dex for \feh~(\citealt{2015RAA....15.1095L}). 
% ==============================================================================
\section{Validation process with the SCR method}
\label{sec:method}

\begin{figure}
\centering
	\scriptsize  
	\tikzstyle{formata}=[rectangle,draw,thin,fill=white,align=center,rounded corners, draw=black, fill=green!30]
	\tikzstyle{formatb}=[rectangle,draw,thin,fill=white,text width=14em,align=center,rounded corners, draw=black, fill=green!30]
	\tikzstyle{formatg}=[rectangle,draw,thin,fill=white, draw=black, fill=yellow!30]
	\tikzstyle{formatf}=[rectangle,draw,thin,fill=white,text width=30em,align=center,rounded corners, draw=black, fill=green!30]
	\tikzstyle{line}=[draw, thick, -latex',shorten >=2pt]
	\tikzstyle{linered}=[draw, thick];\tikzstyle{point}=[coordinate,on grid,]  
	\begin{tikzpicture}
	\node[formata] (a){(a) Combine PS1 DR1 with Gaia EDR3 and LAMOST DR7};
	\node[formatb,below of=a,node distance=10.5mm] (b){(b) Select calibration sample and control sample};
	\node[formatb,below of=b,node distance=14.6mm] (d){(c) Fit the intrinsic colors as functions of ($T_{\rm eff}$, $\rm [Fe/H]$) with initial reddening coefficient $\bf R_{\rm ini}$ for the control sample};
	\node[formatb,below of=d,node distance=16.mm] (newd){(d) Correct for the magnitude-dependent fitting residuals by linear regression};
	\node[formatb,below of=newd,node distance=16.mm] (e){(e) Linear regression of reddening coefficients $\bf R$ with respect to $E(G_{\rm BP}-G_{\rm RP})$ for the calibration sample};
	\node[formatf,below of=e,node distance=14.6mm] (f){(f) Apply reddening correction and magnitude-dependent
	
	correction to the calibration sample, and obtain $\bf \Delta M$(RA, Dec)};
	\node[point,right of=e,node distance=35mm] (h){}; \node[point,below of=d,node distance=5.3mm] (i){};
	\node[point,right of=d,node distance=35mm] (j){};
	\node[formatg,right of=i,node distance=35mm](g){$\bf R$ to $\bf R_{\rm ini}$};
	\node[point,below of=b,node distance=10mm] (m){};\node[point,left of=m,node distance=15mm] (l){};
	\node[point,below of=e,node distance=0.6mm] (n){};\node[point,right of=n,node distance=15mm] (r){};
	\node[point,left of=newd,node distance=9em] (itl){};\node[point,left of=d,node distance=9em] (itu){};
	\draw[line, ->](a)--(b);\draw[line, ->](b)--(d);\draw[line, ->](d)--(newd);\draw[line, ->](newd)--(e);
	\draw[line, ->](e)--(f);\draw[-](e)--(h);\draw[->](h)--(g);\draw[-](g)--(j);\draw[->](j)--(d);
	\node[point,below of=b,node distance=6.3mm] (uc){};\node[point,left of=uc,node distance=9.8em] (ul){};
	\node[point,right of=uc,node distance=43.5mm] (ur){};\node[point,below of=e,node distance=7.8mm] (bc){};
	\node[point,left of=bc,node distance=9.8em] (bl){};\node[point,right of=bc,node distance=7em] (brc){};
	\node[point,right of=brc,node distance=43.5mm-7em] (br){};
	\draw[linered,-,draw=red,dashed](ul)--(ur);\draw[linered,-,draw=red,dashed](bl)--(brc);
	\draw[linered,-,draw=red,dashed](brc)--node[above]{\textcolor{red}{$Iterate$}}(br);
	\draw[linered,-,draw=red,dashed](ul)--(bl);\draw[linered,-,draw=red,dashed](ur)--(br);
	\draw[-](newd)--(itl);\draw[-](itl)--(itu);\draw[->](itu)--(d);
	\end{tikzpicture} 
    \caption{{\small Flowchart of the SCR method in this work for a given PS1 passband.}}
    \label{Fig:flow}
\end{figure}
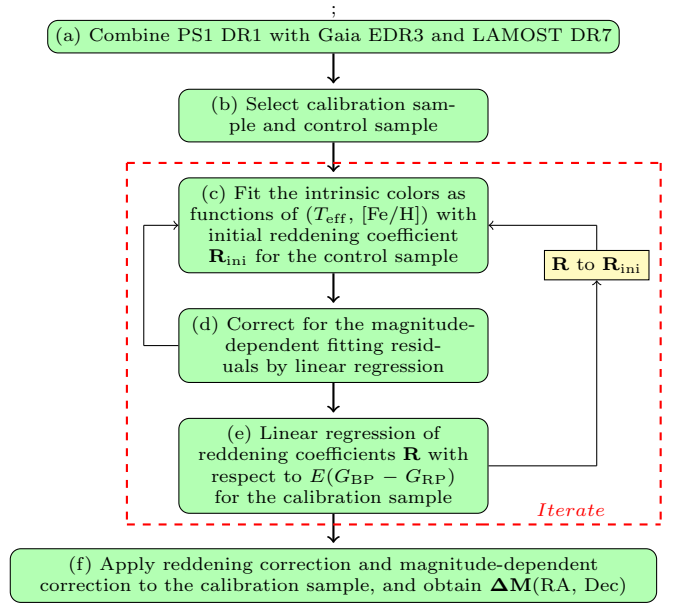
\begin{figure}[ht!]
\resizebox{\hsize}{!}{\includegraphics{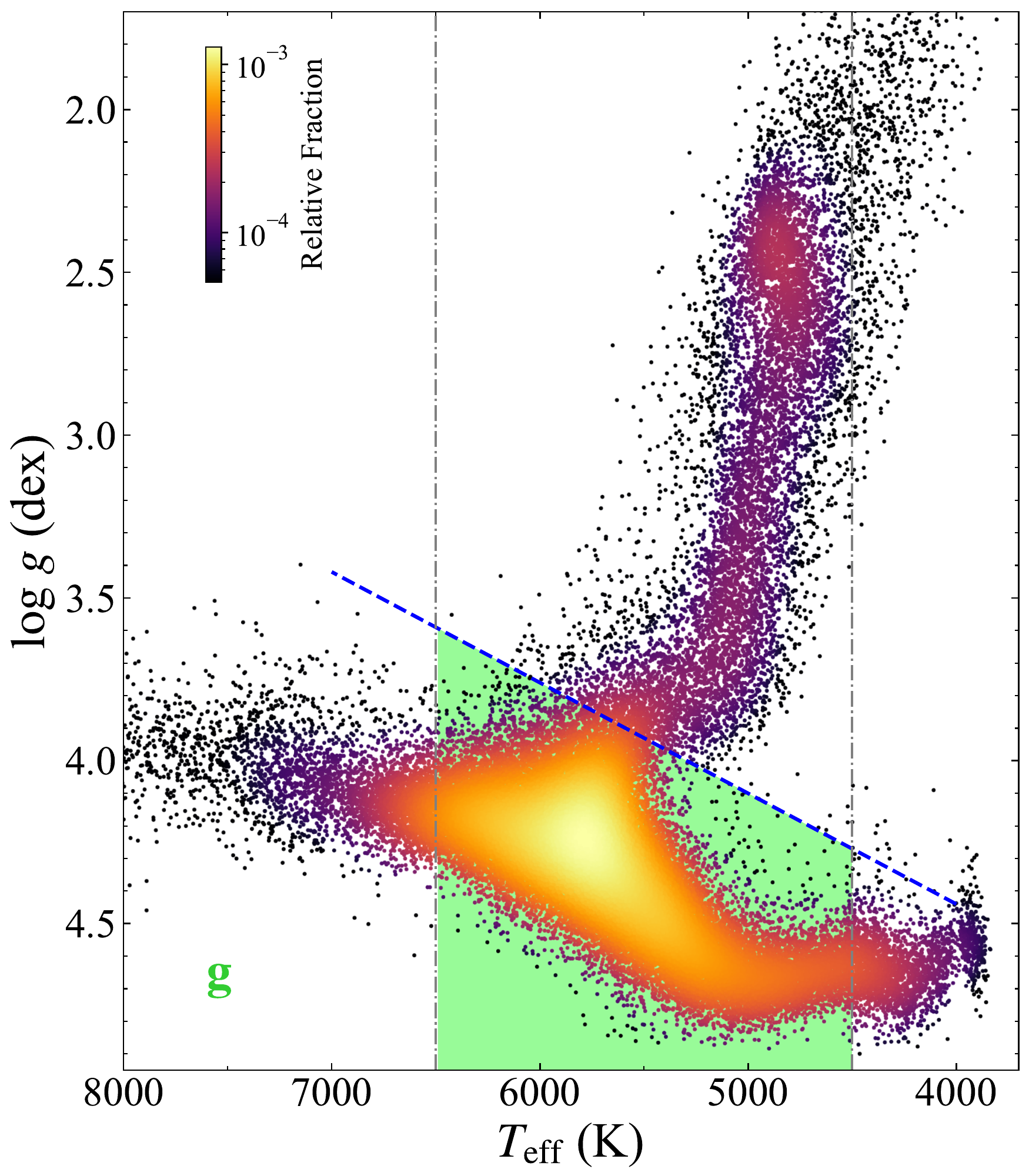}}
\caption{{\small Distribution of stars in the $T_{\rm eff}$-$\log g$ plane for the $g$ band. Calibration sample stars are distributed in the green-background region. The two vertical dotted lines mark $T_{\rm eff}$ = 6500\,K and 4500\,K. The blue dotted line marks the boundary between the main-sequence and giant stars. A color bar is over-plotted at the upper-left corner indicating the normalized stellar densities.}}
\label{Fig:logg}
\end{figure}
\begin{figure*}[ht!]
\resizebox{\hsize}{!}{\includegraphics{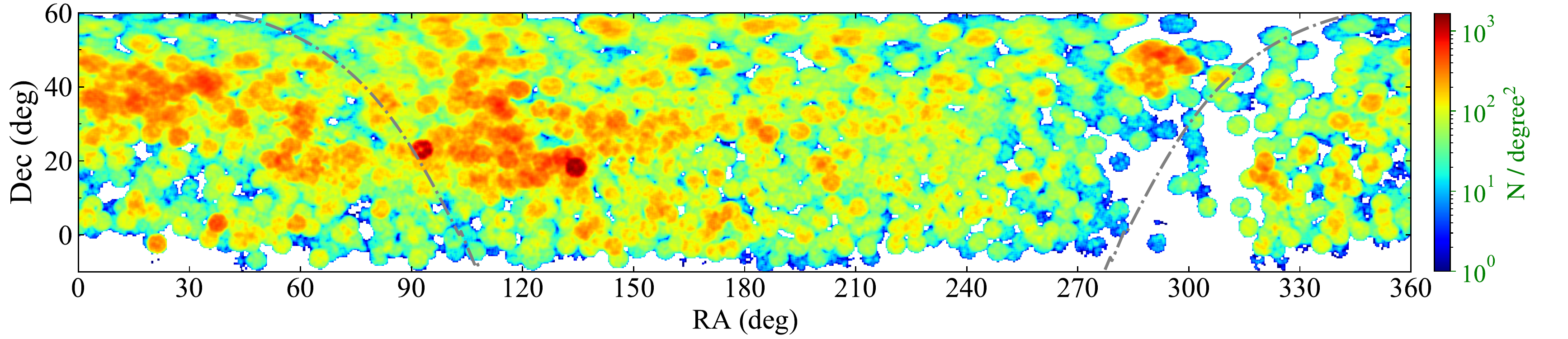}}
\caption{{\small Spatial distribution of the calibration sample stars for the $g$ band. A color bar is over-plotted to the right indicating the stellar number densities. The gray dotted line indicates the Galactic plane.}}
\label{Fig:ra_dec}
\end{figure*}
\begin{figure*}[ht!]
\resizebox{\hsize}{!}{\includegraphics{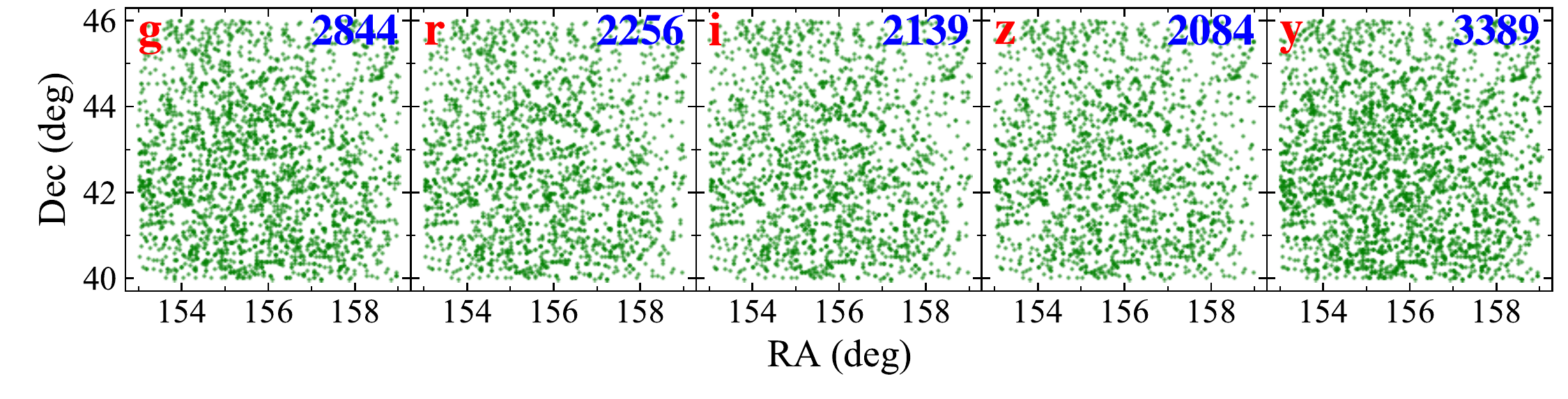}}
\caption{{\small Spatial distributions of the control sample stars for the $grizy$ bands. The band and the star number are labeled in each panel.}}
\label{fig:control}
\end{figure*}

An overview of the validation process using the SCR method to the PS1 calibration 
is shown in Figure\,\ref{Fig:flow}. The details are as below:
\begin{enumerate}
  \item[a.] Combine the PS1 DR1 photometric data with the Gaia EDR3 and the LAMOST DR7. 
  The adopted cross-matching radius is $1''$.
  \item[b.] Select main sequence stars (${\log g}>-3.4\times 10^{-4}\times T_{\rm eff}+5.8$) as the calibration samples with the following constraints: 1) mag\{$g, r, i, z$\} $>$ 14 and mag\{$y$\} $>$ 13 to avoid saturation; 2) error\{$g, r, i, z, y$\} $<$ 0.02 mag; 
  3) \texttt{phot}\_\texttt{bp}\_\texttt{rp}\_\texttt{excess}\_\texttt{factor} $<$ $1.3+0.06\times(G_{\rm BP}-G_{\rm RP})^2$ to avoid bad Gaia photometry; 4) $4500<T_{\rm eff}<6500$\,K, a relatively narrow temperature range for robust fitting of intrinsic colors with atmospheric parameters and the temperature-dependence reddening coefficients, but still with sufficient numbers of stars; 5) $\rm [Fe/H]>-1$ for an easy intrinsic color fitting with atmospheric parameters; and 6) Signal-to-noise ratio for the $g$ band ($SNR_{\rm g}$) of the LAMOST spectra $>20$. Finally, 1,688,097, 1,299,006, 1,162,825, 1,113,294 and 1,931,225 stars are selected in the $g$, $r$, $i$, $z$ and $y$ bands, respectively. The calibration sample in the $g$ band is shown in Figure\,\ref{Fig:logg} and \ref{Fig:ra_dec}. Then, the control sample stars are selected as those within a small and low-extinction area ($153<\rm {RA}<159~\rm {degree},~40<Dec<46~\rm {degree}$). % and of low extinction ($E(G_{\rm BP}-G_{\rm RP}) < 0.03$).
  A total number of 2844, 2256, 2139, 2084 and 3389 control stars are selected for the $grizy$ bands, respectively.
  Their spatial distributions are shown in Figure\,\ref{fig:control}.
  
  For reddening correction, the dust reddening map of \citet{1998ApJ...500..525S} is not used 
  as it fails at low Galactic latitudes and shows spatially-dependent systematic errors (Sun et al., submitted). In this work, the values of $E(G_{\rm BP}-G_{\rm RP})$ obtained with the star-pair method \citep{2013MNRAS.430.2188Y,2020ApJ...905L..20R} are adopted instead. 

  \item[c.] Five colors $\bf C={\bf G}_{\rm BP,RP}-{\bf M}^{\rm obs}$ are adopted for the $grizy$ bands, where ${\bf G}_{\rm BP,RP}=(\begin{array}{c} G_{\rm BP},~G_{\rm RP},~G_{\rm RP},~G_{\rm RP},~G_{\rm RP} \end{array})^{\mathrm T}$ and ${{\bf M}^{\rm obs}}=(\begin{array}{c} g,~r,~i,~z,~y \end{array})^{\mathrm T}$. Then, for the control sample, a 2nd-order two-dimensional polynomial (with 6 free parameters, see Equation\,(\ref{intrinsic_color_mod})) as a function of $T_{\rm eff}$ and $\rm [Fe/H]$ is used to fit the intrinsic colors (${\bf C_{\rm 0}}$). Here the intrinsic colors are estimated using Equation\,(\ref{intrinsic_color_obs}), where $\bf R$ represents reddening coefficients.

  \begin{eqnarray}
  \begin{aligned}
  {\bf C}^{\rm mod}_{\rm 0}=~&{\bf a_0}\cdot T_{\rm eff}^2+{\bf a_1}\cdot {\rm [Fe/H]}^2+\\ &{\bf a_2}\cdot T_{\rm eff}\cdot {\rm [Fe/H]}+\\ &{\bf a_3}\cdot T_{\rm eff}+{\bf a_4}\cdot {\rm [Fe/H]}+{\bf a_5}~,  \label{intrinsic_color_mod}
  \end{aligned}
  \end{eqnarray}
\begin{eqnarray}
  {\bf C_{\rm 0}}={\bf C} -{\bf R} \times E(G_{\rm BP}-G_{\rm RP})~,  \label{intrinsic_color_obs}
  \end{eqnarray}
  where
 \begin{equation}\nonumber
 {\bf C}=\left( \begin{array}{c}G_{\rm BP}-g\\G_{\rm RP}-r\\G_{\rm RP}-i\\G_{\rm RP}-z\\G_{\rm RP}-y \end{array} \right),~ 
 {\bf R}=\left( \begin{array}{c}R_{\rm (G_{BP}-g)}\\R_{\rm (G_{RP}-r)}\\R_{\rm (G_{RP}-i)}\\R_{\rm (G_{RP}-z)}\\R_{\rm (G_{RP}-y)} \end{array} \right).
 \end{equation}
 
 %{\bf {An initial set of reddening coefficient values of $(\begin{array}{c} 1,~1,~1,~1,~1 \end{array})^{\mathrm T}$ are given for estimate ${\bf C_{\rm 0}}$ at the first iteration.}}
 
  \item[d.] Moderate magnitude-dependent residuals are found when fitting intrinsic colors 
  as a function of $T_{\rm eff}$ and $\rm [Fe/H]$ of the control samples in each band. 
  Therefore, we use a linear polynomial, ${\bf \Delta M}{\rm (Mag)}=b_{\rm 1}\cdot{\bf M}^{\rm obs}+b_{\rm 0}$, to fit the residuals (${\bf C}_{\rm 0}-{\bf C}_{\rm 0}^{\rm mod}(T_{\rm eff},~\rm [Fe/H])$). 
  To account for magnitude-dependent errors in the PS1 data, corrected magnitudes 
  are obtained by ${\bf M}^{\rm obs}+{\bf \Delta M}{\rm (Mag)}$. Then, we put the corrected magnitudes into the previous step (Figure\,\ref{Fig:flow}(c)). Iterations are performed. 
  
  \item[e.] Based on the results from the above two steps, 
  intrinsic colors ${\bf C_{\rm 0}^{\rm mod}}(T_{\rm eff},~\rm [Fe/H])$ are obtained for the calibration stars, so are their reddening values $\bf C-C_{\rm 0}^{\rm mod}$. Then, the reddening coefficients $\bf R$ with respect to $E(G_{\rm BP}-G_{\rm RP})$ are derived by linear regression with $3\sigma$ clipping.
  Note that the lines are not forced to pass through the origin. The non-zero offsets, ${\bf \delta M}$, represent the zero-point differences between the control samples and 
  the calibration samples. Iterations are also needed here, as shown in Figure\,\ref{Fig:flow}. Finally, ${\bf \delta M}=(+0.0047,-0.0027,-0.0008,+0.0061,-0.0013)^{\mathrm T}$ mag. 
  
  In this process, because of the very broad passbands of $G_{\rm BP}$ and $G_{\rm RP}$, 
  we have also considered the influence of temperature on the reddening coefficients. 
  Temperature-dependent reddening coefficients are adopted
  for the $G_{\rm BP} - g$ and $G_{\rm RP} - r$ colors. 
  
  \item[f.] At last, 
  the predicted model magnitudes ${\bf M^{\rm mod}}$ and magnitude offsets ${\bf \Delta M}({\rm RA, Dec})$ can be obtained from Equations\,(\ref{e2}) and (\ref{e3}):
  \begin{eqnarray}
    \begin{aligned}
    {\bf M^{\rm mod}}=~&{\bf G}_{\rm BP,RP}-{\bf C_{\rm 0}^{\rm mod}}(T_{\rm eff},~\rm [Fe/H])-\\
    &{\bf R} \times E(G_{\rm BP}-G_{\rm RP})~, \label{e2}
    \end{aligned}
    \end{eqnarray}
    \begin{eqnarray}
    {\bf \Delta M}({\rm RA, Dec})={\bf M^{\rm mod}}-{\bf M^{\rm obs}}-{\bf \delta M}~.  \label{e3}
    \end{eqnarray}    
\end{enumerate}
  
\begin{figure*}[ht!]
\resizebox{\hsize}{!}{\includegraphics{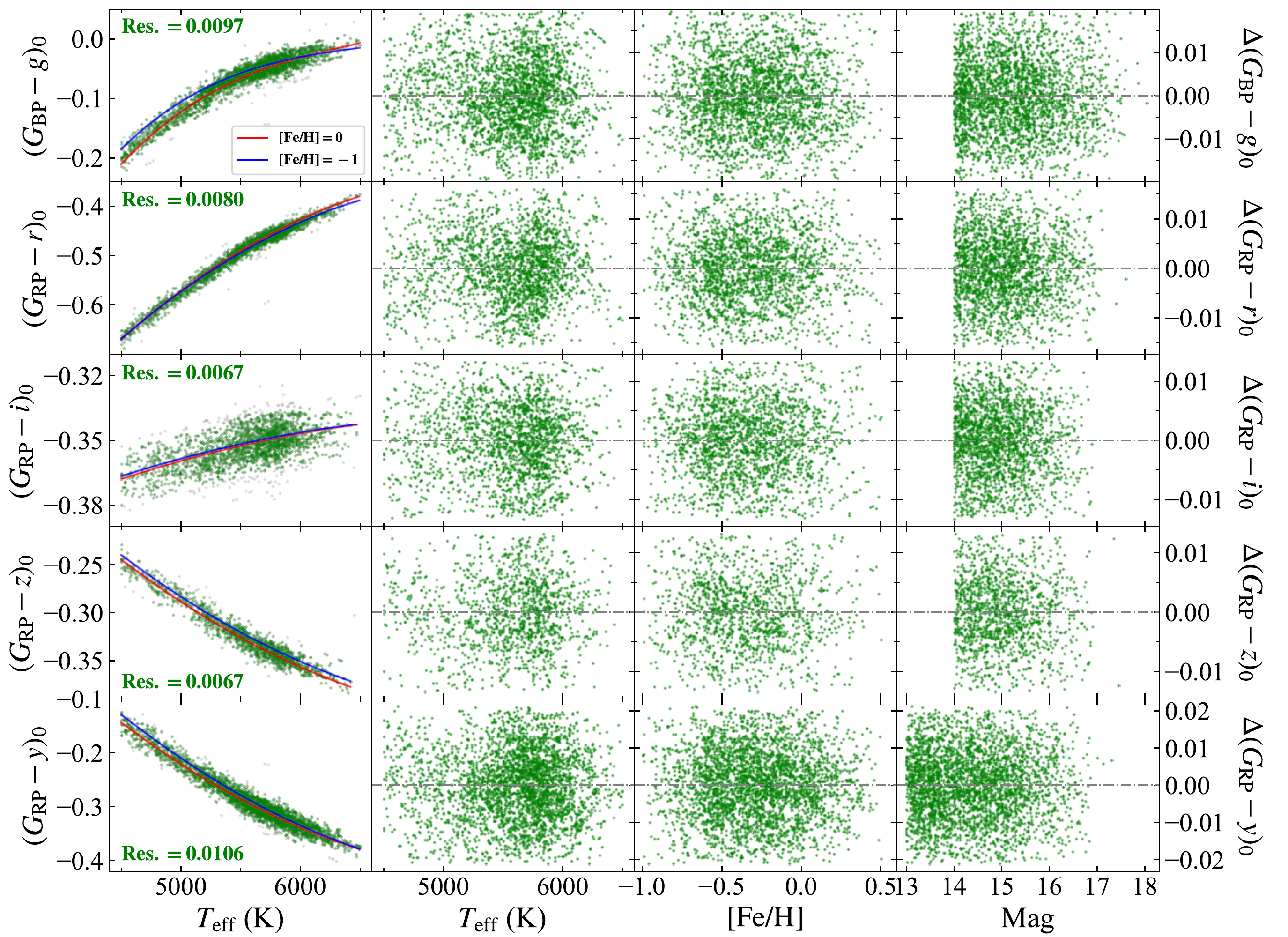}}
\caption{{\small Two-dimensional 2-order polynomial fitting (with 6 free parameters) of intrinsic colors as functions of \teff~and \feh for the control stars. From top to bottom are for the \bpg, \rpr, \rpi, \rpz~and \rpy~colors, respectively. The left column shows the fitting results after $3\sigma$ clipping, and the outliers are indicated by gray pluses. The red and blue curves represent results for $\rm [Fe/H]$ = 0 and $-$1, respectively, and the fitting residuals are labeled. The 2nd, 3rd and 4th columns plot residuals against \teff, \feh~ and magnitude, respectively.}}
\label{Fig:fitting}
\end{figure*}
\begin{deluxetable*}{ccccccc}[ht!]
\tablecaption{The coefficients used to obtain intrinsic colors as functions of \teff~and \feh~ in the five bands. \label{tab:1}}
\tablehead{
\colhead{Intrinsic Color} & \colhead{$T_{\rm eff}^2$} & \colhead{${\rm [Fe/H]}^2$} & \colhead{$T_{\rm eff}\cdot{\rm [Fe/H]}$} & \colhead{$T_{\rm eff}$} & \colhead{$\rm [Fe/H]$} & \colhead{Constant Term}}
\startdata
$(G_{\rm BP}-g)_{\rm 0}$ & $-4.520\times 10^{-8}$ & $-5.120\times 10^{-3}$ & $~~\,1.723\times 10^{-5}$ & $~~\,5.892\times 10^{-4}$ & $-1.097\times 10^{-1}$ & $-1.940$ \\
$(G_{\rm RP}-r)_{\rm 0}$ & $-3.407\times 10^{-8}$ & $~~\,4.415\times 10^{-3}$ & $~~\,4.052\times 10^{-6}$ & $~~\,5.192\times 10^{-4}$ & $-1.456\times 10^{-2}$ & $-2.317$ \\
$(G_{\rm RP}-i)_{\rm 0}$ & $-2.742\times 10^{-9}$ & $~~\,3.609\times 10^{-3}$ & $~~\,5.390\times 10^{-7}$ & $~~\,4.278\times 10^{-5}$ & $~~\,7.821\times 10^{-5}$ & $-0.505$ \\
$(G_{\rm RP}-z)_{\rm 0}$ & $~~\,1.314\times 10^{-8}$ & $~~\,1.310\times 10^{-3}$ & $-3.992\times 10^{-7}$ & $-2.124\times 10^{-4}$ & $-1.899\times 10^{-3}$ & $~~\,0.444$ \\
$(G_{\rm RP}-y)_{\rm 0}$ & $~~\,2.462\times 10^{-8}$ & $-5.528\times 10^{-3}$ & $~~\,7.085\times 10^{-6}$ & $-3.881\times 10^{-4}$ & $-5.183\times 10^{-2}$ & $~~\,1.104$
\enddata
\end{deluxetable*}
\begin{figure*}[ht!]
   \centering
  \resizebox{\hsize}{!}{\includegraphics{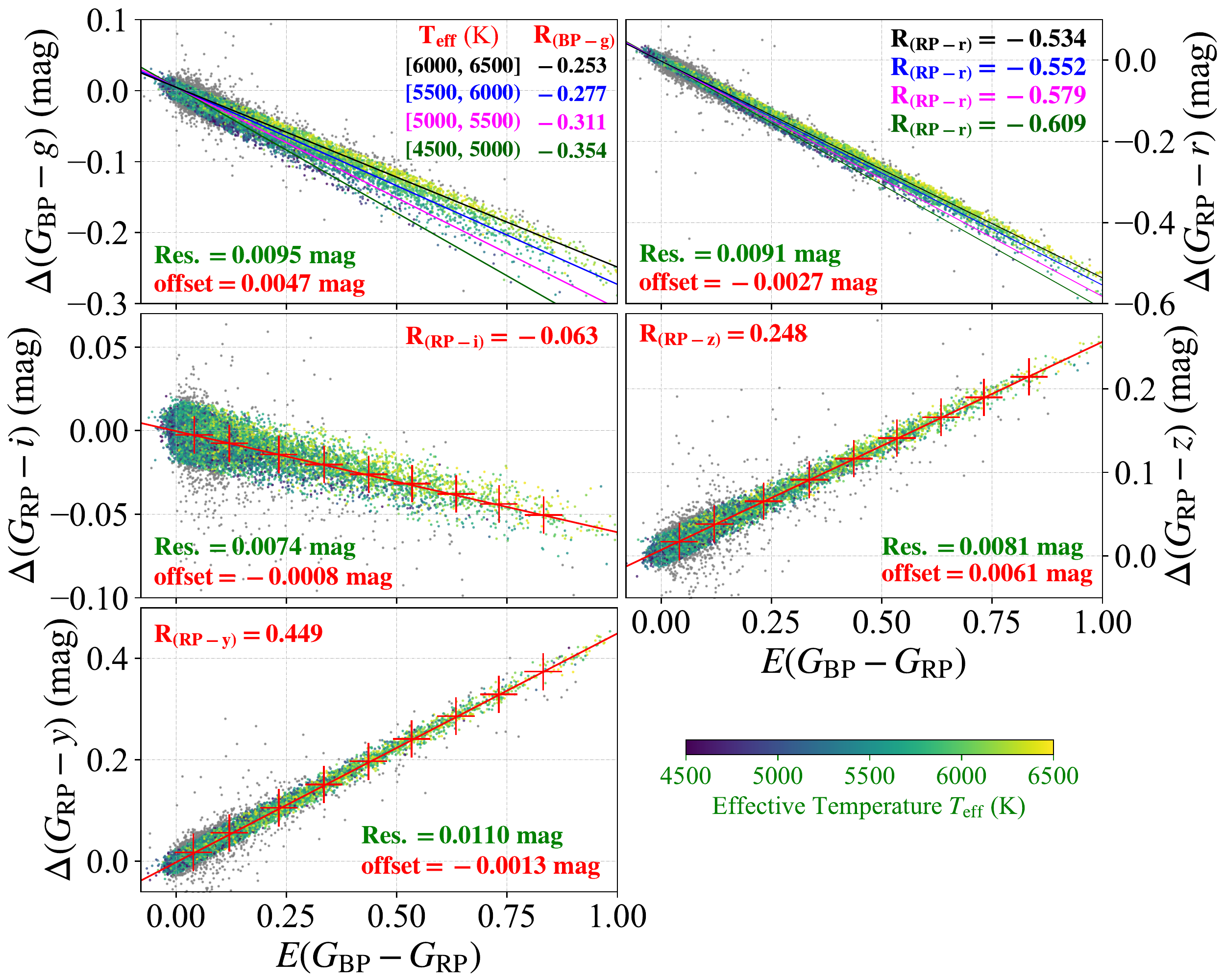}}
   \caption{{\small Linear regression of reddening coefficients of the \bpg, \rpr, \rpi, \rpz~and \rpy~colors with respect to \ebprp~for the calibration samples. To avoid crowdness, 
   only one in fifty stars are plotted. 
   Colors indicate different temperatures. 
   For the middle and bottom panels where reddening coefficients show weak dependence on temperature, 
   the red pluses are the median values after $3\sigma$ clipping and the outliers are indicated by gray pluses; the red lines are linear fits to the red pluses; the slopes, offsets, and fitting residuals are all labeled. 
   For the two top panels where reddening coefficients show moderate temperature dependence, the fitted reddening coefficients for four temperature ranges are shown.
   }}
  \label{Fig:regc}
\end{figure*}
\begin{deluxetable*}{cccccc}[ht!]
\tablecaption{Temperature-dependent reddening coefficients of \bpg~and \rpr~with respect to \ebprp. \label{tab:2}}
\tablehead{
\colhead{Color} & \colhead{$T_{\rm eff}^3$} & \colhead{$T_{\rm eff}^2$} & \colhead{$T_{\rm eff}$} & \colhead{Constant}}
\startdata
$G_{\rm BP}-g$  & $-1.765\times 10^{-12}$ & $1.094\times 10^{-8}$ & $~~\,1.080\times 10^{-4}$ & $-0.925$ \\
$G_{\rm RP}-r$  & $-1.434\times 10^{-11}$ & $2.183\times 10^{-7}$ & $-1.041\times 10^{-3}$ & $~~\,0.941$ \\
\enddata
\end{deluxetable*}
\begin{figure*}[ht!]
   \centering
   \includegraphics[width=15.cm]{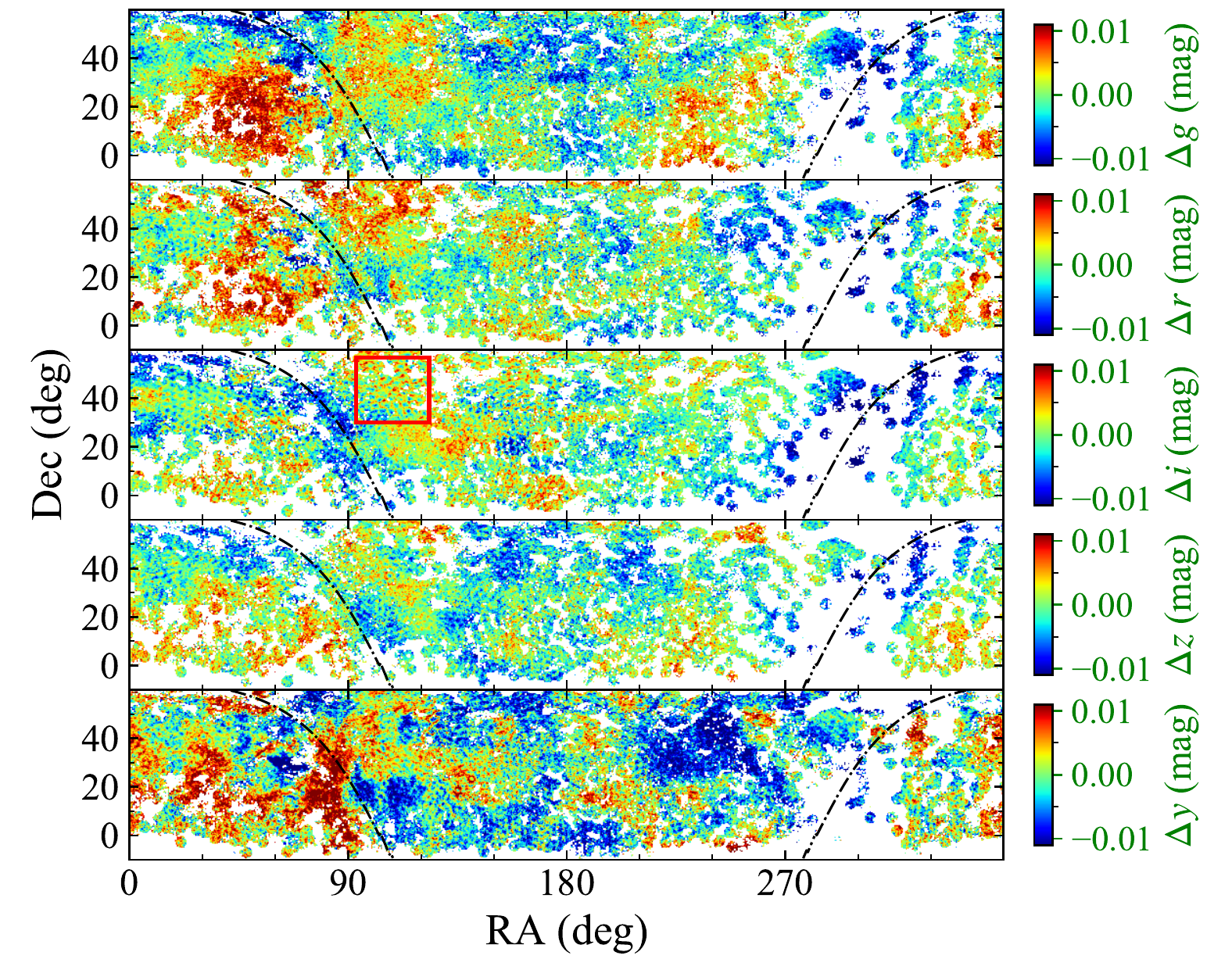}
   \caption{{\small Spatial variations of magnitude offsets after a $20' \times 20'$ binning in the $g$, $r$, $i$, $z$ and $y$ bands. The black dotted line indicates the Galactic plane in each panel. 
   Color bars are over-plotted to the right. The small region marked with a red box in the $i$ band is re-plotted in Figure\,\ref{Fig:delmag_redec_sub}.}}
  \label{Fig:delmag_radec_ini}
\end{figure*}
\begin{figure}[ht!]
   \centering
   \resizebox{\hsize}{!}{\includegraphics{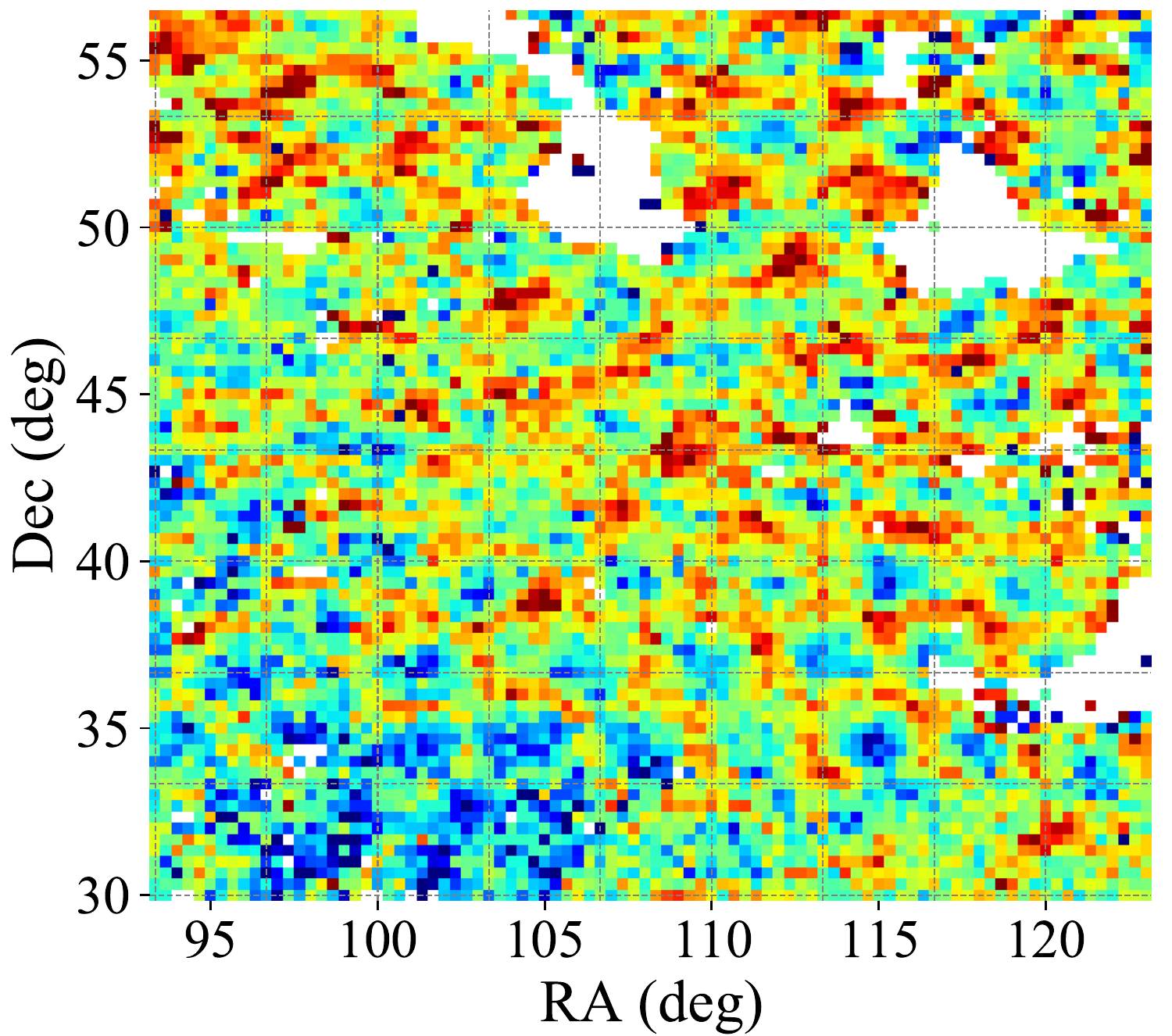}}
   \caption{{\small A zoom in plot for the small region in Figure\,\ref{Fig:delmag_radec_ini} (red box) in the $i$ band with the same color bar. The spacing of the gray grids is $3.3^{\circ}$. 
   }}
  \label{Fig:delmag_redec_sub}
\end{figure}
\begin{deluxetable}{ccc}[ht!]
\tablecaption{The coefficients used for magnitude-dependent corrections. \label{tab:3}}
\tablehead{
\colhead{${\bf \Delta M} ~{\rm (Mag)}$} & \colhead{Slope ($b_1$)} & \colhead{Constant ($b_0$)}}
\startdata
$\Delta g$ & $0.0050$ & $-0.0755$ \\
$\Delta r$ & $0.0048$ & $-0.0727$ \\
$\Delta i$ & $0.0045$ & $-0.0686$ \\
$\Delta z$ & $0.0038$ & $-0.0576$ \\
$\Delta y$ & $0.0029$ & $-0.0415$
\enddata
\end{deluxetable}
\begin{figure*}[ht!]
   \centering
   \includegraphics[width=10.cm]{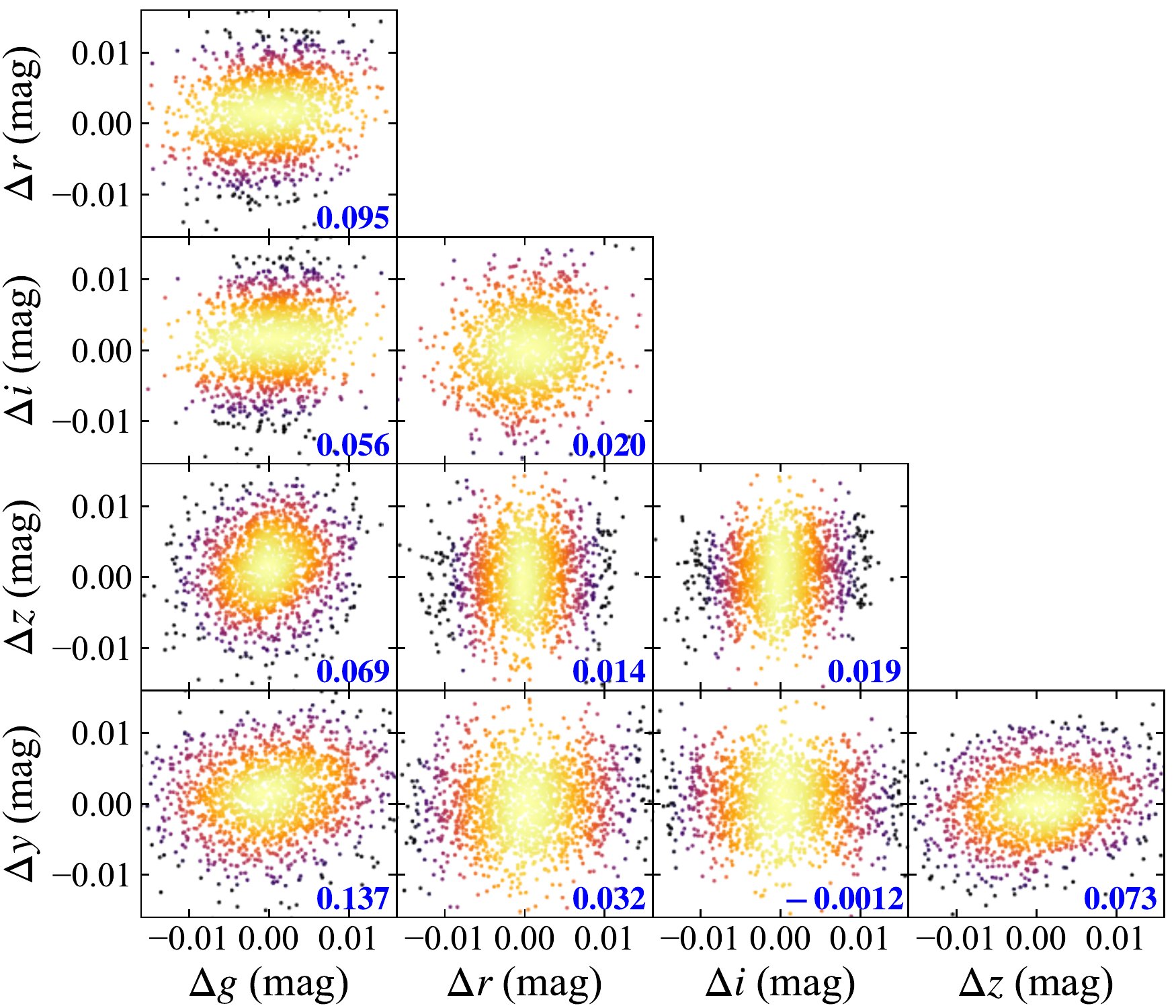}
   %\resizebox{\hsize}{!}{\includegraphics{pearson_R.pdf}}
   \caption{{\small The correlation plots between the magnitude offsets with a restriction of the star numbers of box is more than 5 for each two bands. For each panel, the correlation coefficient are marked. The color in each panel indicates number density of stars.}}
  \label{Fig:R}
\end{figure*} 
\begin{figure*}[ht!]
   \centering
   \includegraphics[width=15.0cm]{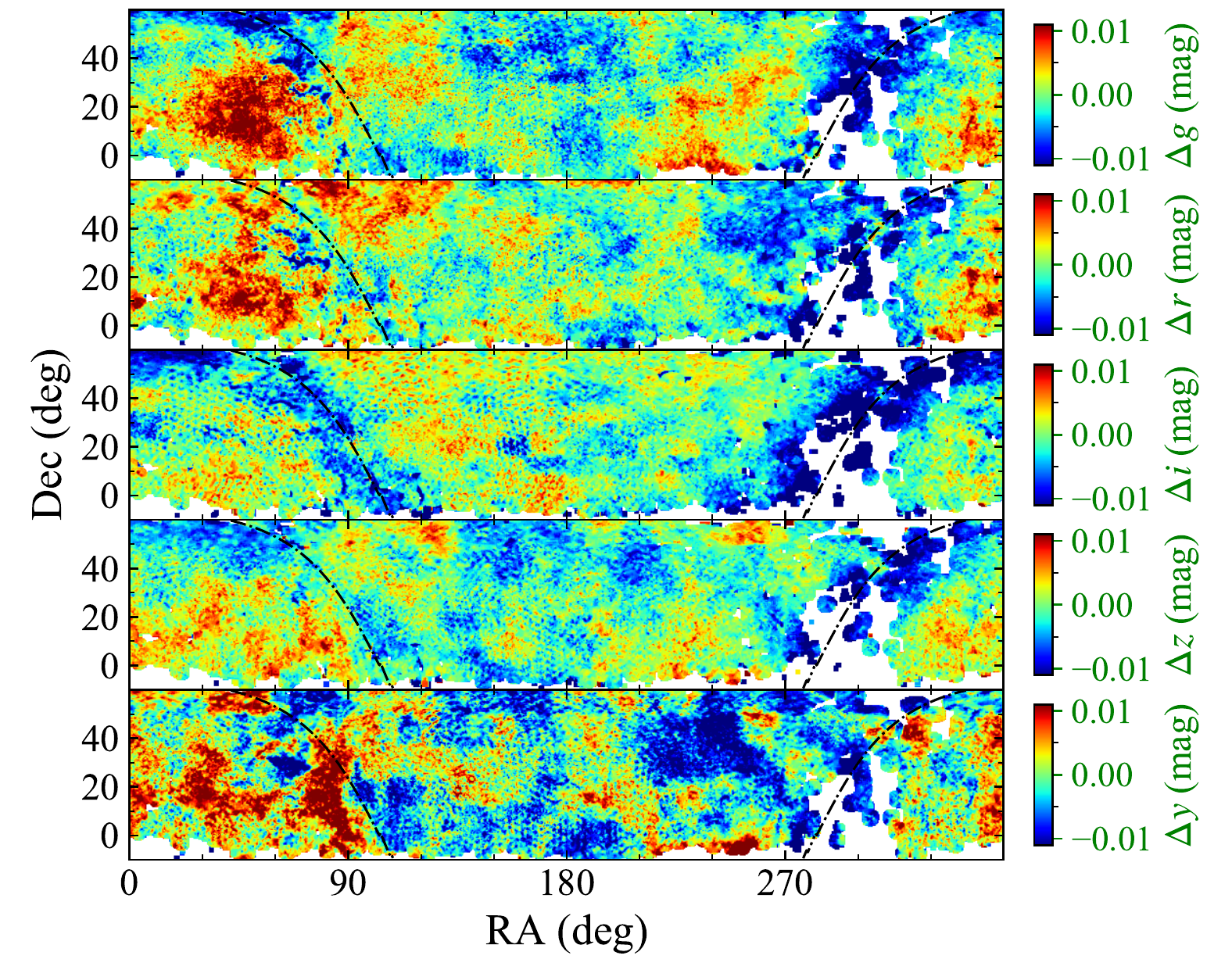}
   %\resizebox{\hsize}{!}{\includegraphics{delmag_radec_c.pdf}}
   \caption{{\small Same to Figure\,\ref{Fig:delmag_radec_ini} but after smoothing.}}
  \label{Fig:delmag_radec_c}
\end{figure*}
\begin{figure*}[ht!]
   \resizebox{\hsize}{!}{\includegraphics{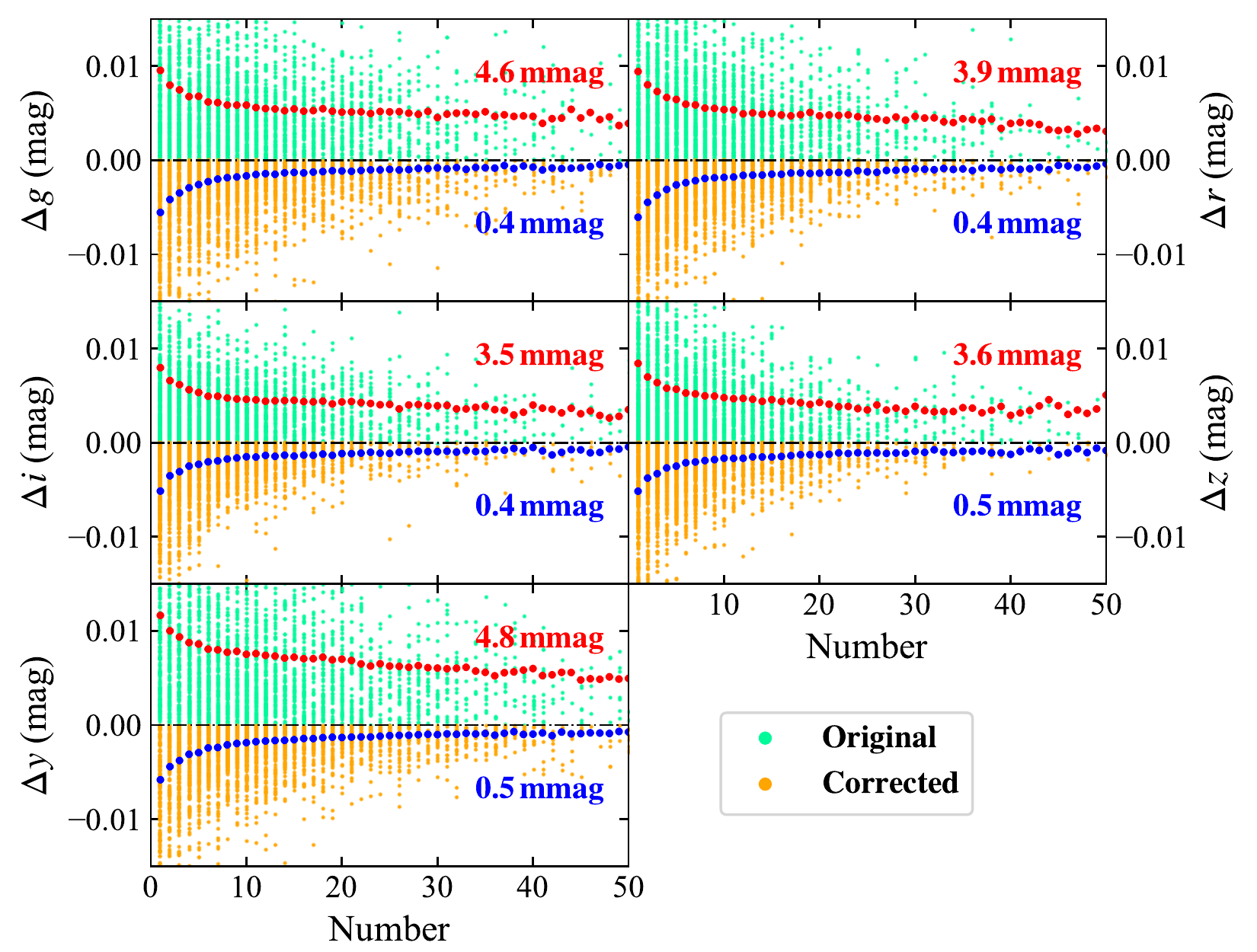}}
   \caption{{\small Magnitude offsets as a function of star numbers in $20' \times 20'$ region before and after correction. Form top to bottom are for the $g$, $r$, $i$, $z$ and $y$ bands, respectively. The green (orange) pluses are the initial (corrected) magnitude offsets, and their standard deviations are indicated by red (blue) dots. The convergence value of the scatter is marked in each panel.}}
  \label{Fig:conv}
\end{figure*} 

% ==============================================================
\section{Results} \label{sec:result}
%\subsection{Spatial variations of magnitude offsets}
\begin{figure}[ht!]
   \resizebox{\hsize}{!}{\includegraphics{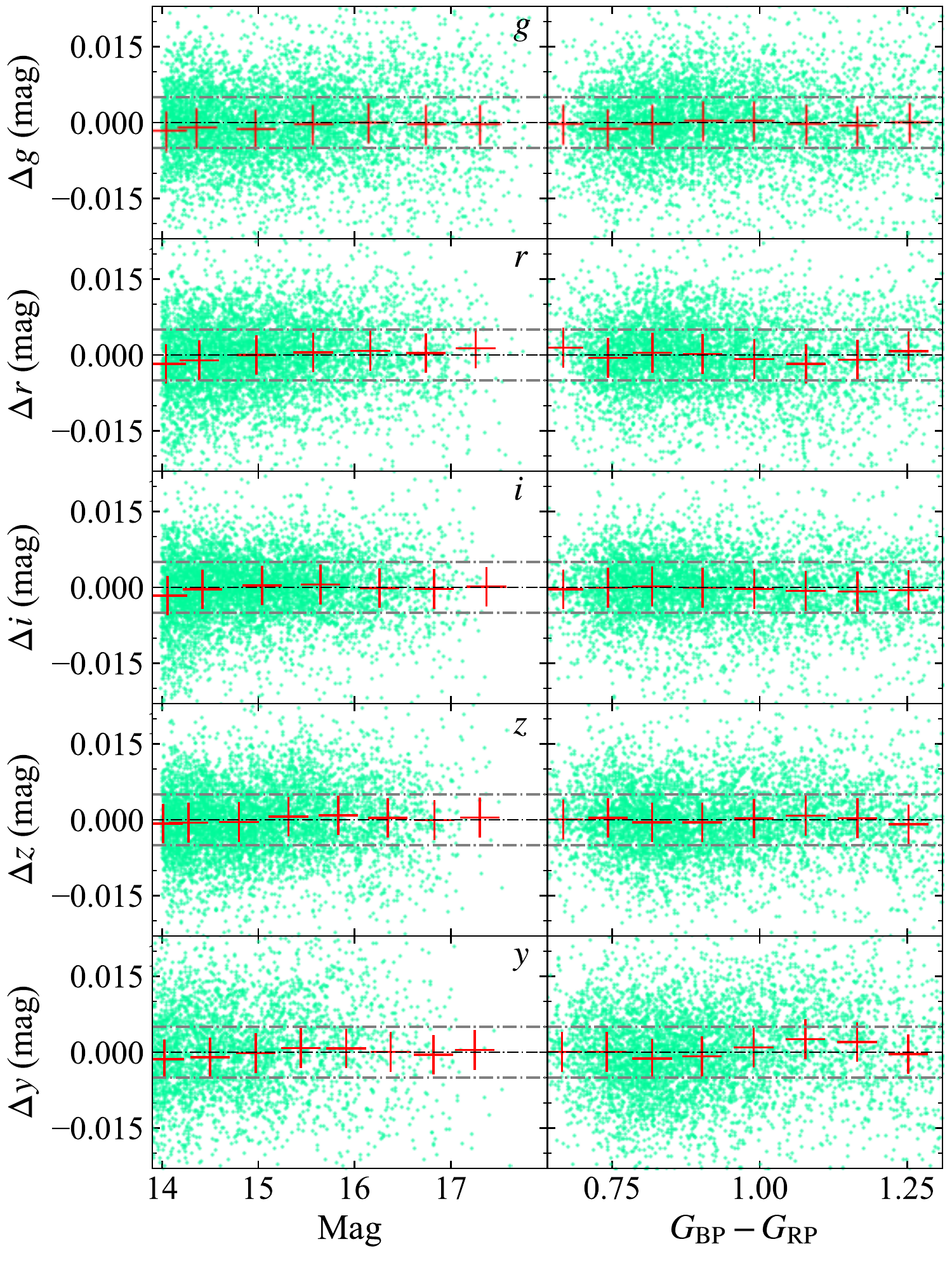}}
   \caption{{\small The variations of magnitude offset as a function of magnitude (left) and \bprp~(right) in $g$, $r$, $i$, $z$ and $y$ bands. The red pluses are the median values.
   The black and gray dotted lines in each panel denote magnitude offsets of 0 and $\pm 5$ mmag, respectively.}}
  \label{Fig:delmag_paras}
\end{figure} 
\begin{figure}[ht!]
   \resizebox{\hsize}{!}{\includegraphics{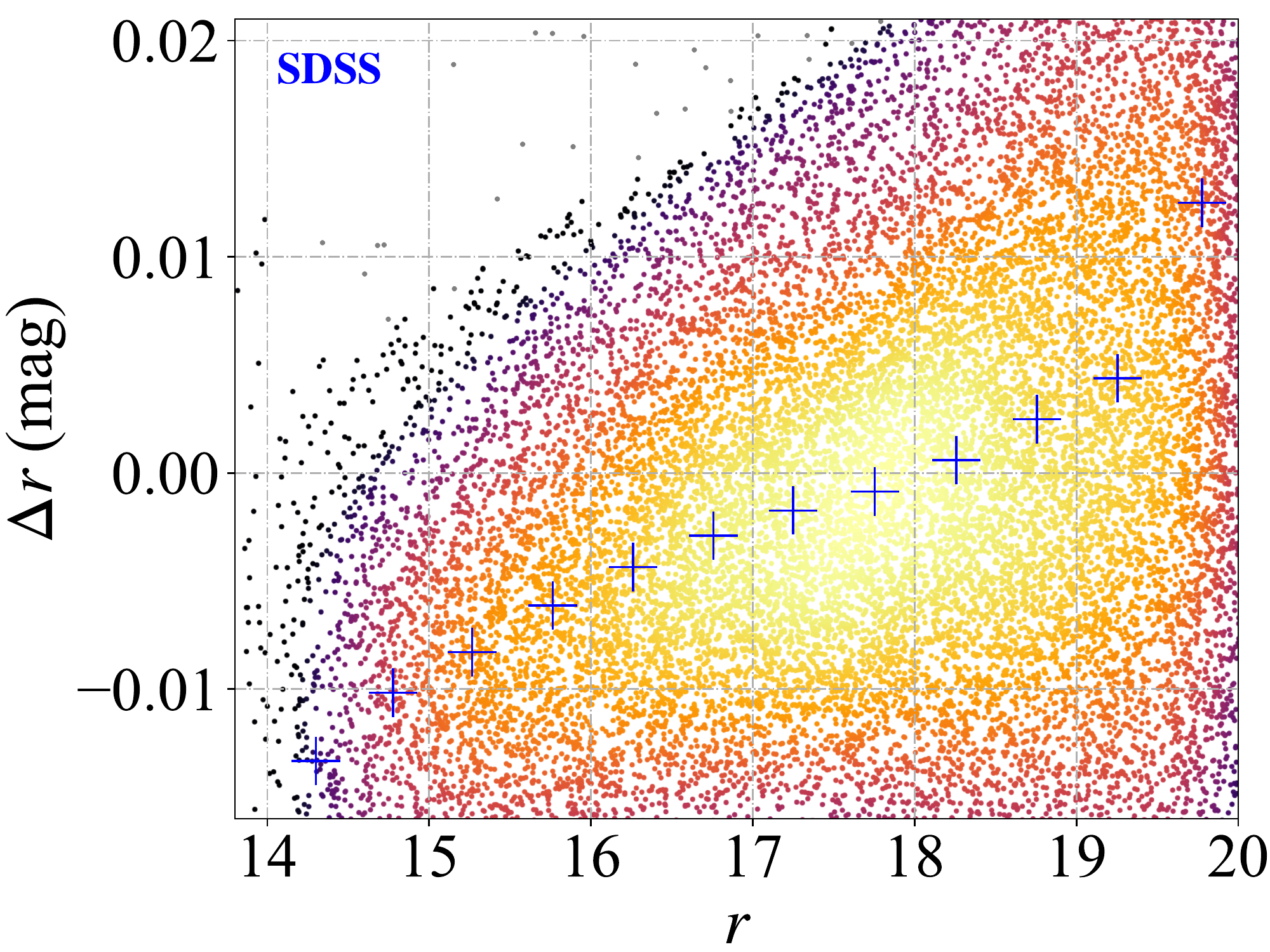}}
   \caption{{\small An example showing the variations of $r$ magnitude offset as a function of $r$ magnitude obtained by comparing with the SDSS stripe\,82 data. The blue pluses are the median
   values. The color represents the number density of stars.}}
  \label{Fig:slr}
\end{figure} 
\begin{figure}[ht!]
   \resizebox{\hsize}{!}{\includegraphics{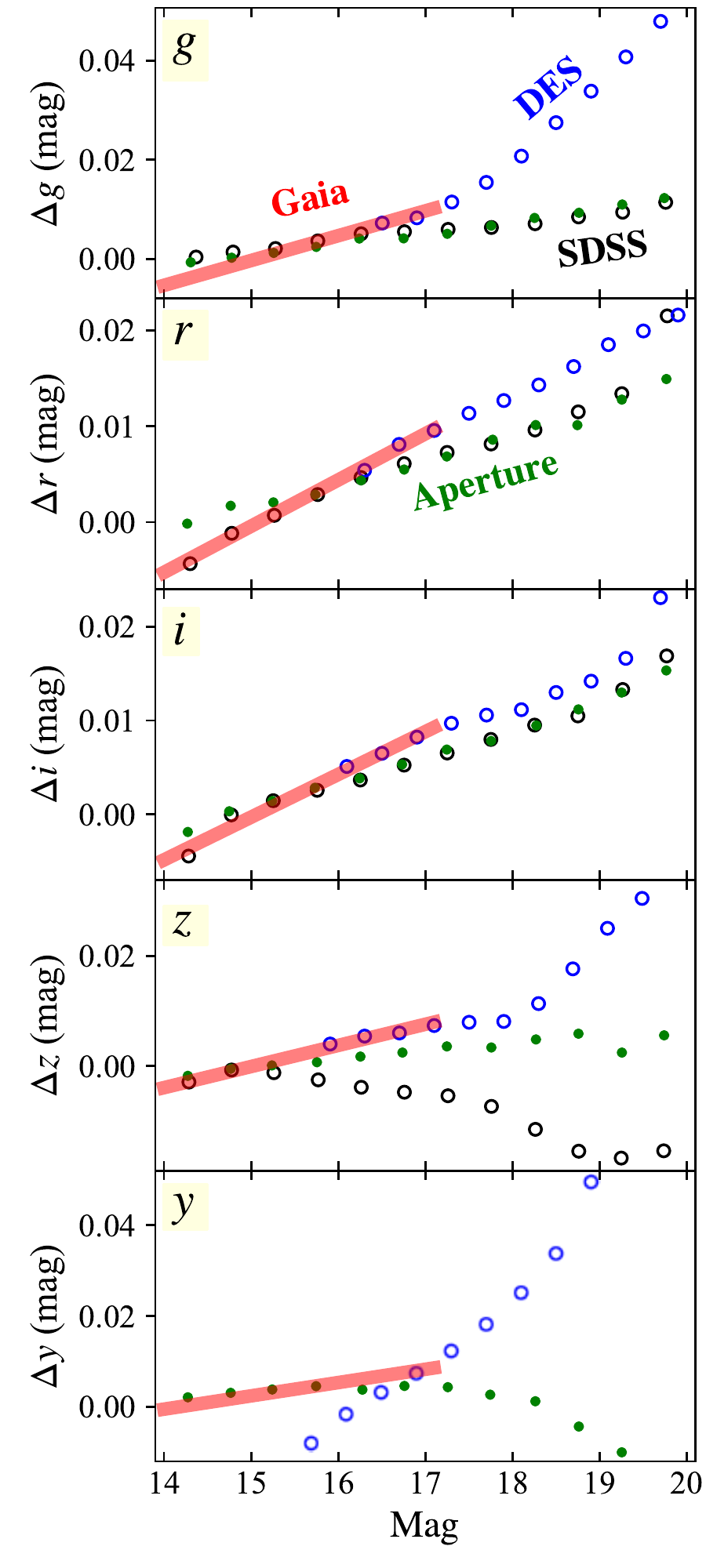}}
   \caption{{\small The variations of magnitude offset as a function of magnitude in the five bands. 
   The black and blue circles denote results from the SDSS Stripe\,82 and DES DR1, respectively. 
   The green dots denote results of comparison with the aperture-based magnitudes of PS1. 
   The red lines denote results from Gaia.}}
  \label{Fig:del_mag2}
\end{figure} 
\begin{figure*}[ht!]
   \resizebox{\hsize}{!}{\includegraphics{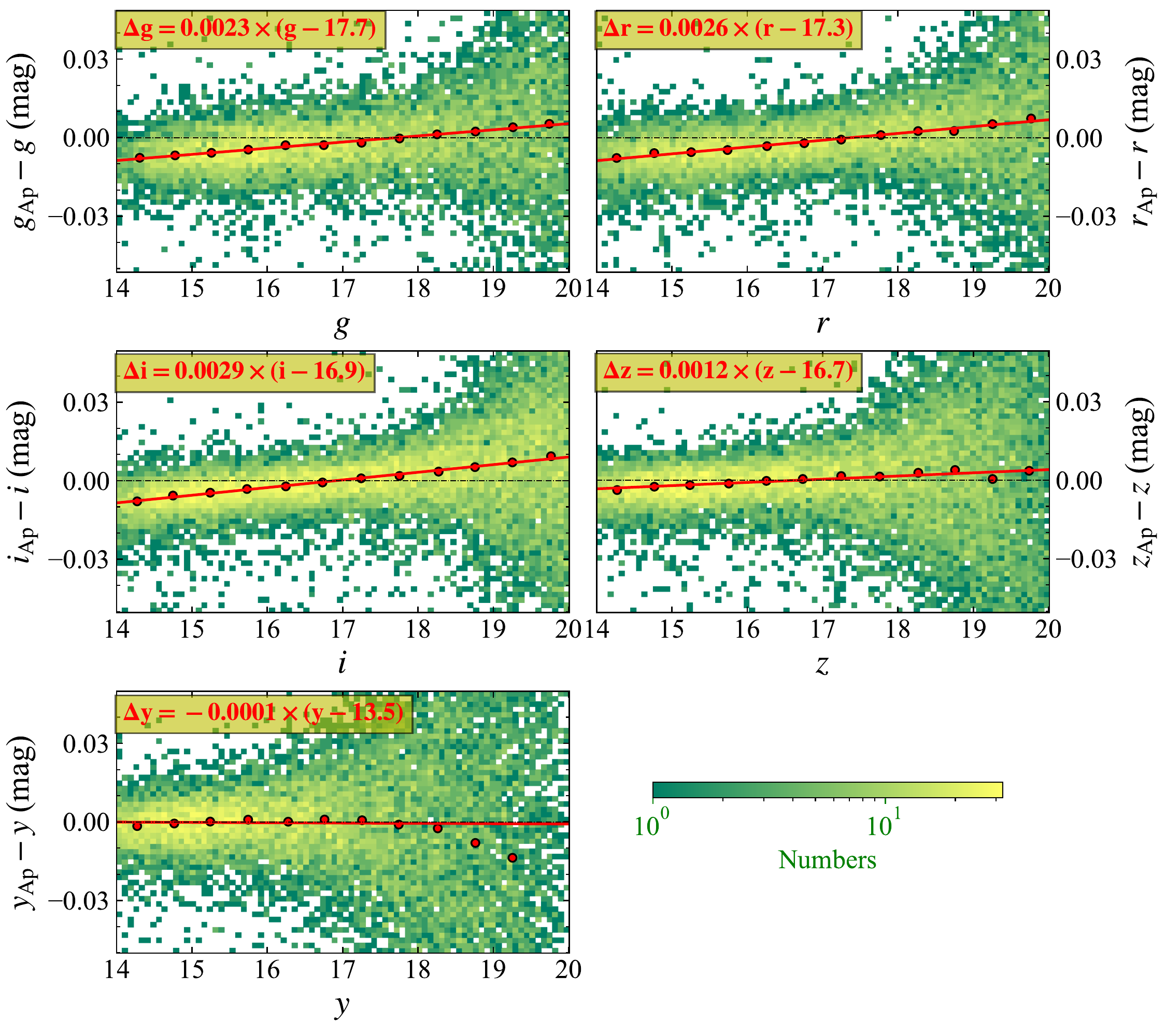}}
   \caption{{\small The difference between the aperture-based and PSF magnitudes as a function of PSF magnitude in the five bands. 
   The black points in each panel are the median values. The linear fitting results to these dots
   are over-plotted in red lines and marked.}}
  \label{Fig:magap_mag}
\end{figure*} 

The final fitting results of the intrinsic colors as a function of (\teff, \feh) are shown in 
Figure\,\ref{Fig:fitting}. The corresponding fitting parameters are listed in Table\,\ref{tab:1}. 
The fitting residuals are respectively 0.0097, 0.0080, 0.0067, 0.0067 and 0.0106 mag for the \bpg, \rpr, \rpi, \rpy~and \rpz~colors,
suggesting that one can predict PS1 magnitudes to a precision of 1 per cent or better from the LAMOST 
and Gaia data for individual stars. 
The fitting residuals show no dependence on \teff~and \feh, and magnitude either after 
correcting the magnitude-dependent errors.
The final fitting coefficients of magnitude-dependence corrections ${\bf \Delta M}\rm (Mag)$ are listed in Table\,\ref{tab:3}. The slopes imply that magnitude-dependent corrections are 0.005, 0.005, 0.005, 0.004, 0.003 mag per magnitude in the 14 -- 17 magnitude range for the $grizy$ bands, respectively. Note that the magnitude-dependent corrections are relative. The corrections are zero at ($g, r, i, z, y$) = (15.1, 15.1, 15.2, 15.2, 14.3), which are the typical magnitudes of the control samples. 

The results of reddening coefficients with respect to \ebprp~in the five colors 
are plotted in Figure\,\ref{Fig:regc}. 
It can be seen that the reddening coefficients show moderate dependence on stellar temperature for 
the \bpg~and \rpr~colors. Therefore, taking 500\,K as the bin width, the calibration stars are divided into 16 overlapping bins according to their temperature. Reddening coefficients of the 16 bins are
further obtained via linear regression, with the same offset. 
Then, we use a cubic polynomial of 4 free parameters to fit the reddening coefficients 
as a function of \teff. The final fitting parameters are given in Table\,\ref{tab:2}. 
The temperature dependence of reddening coefficients for the \rpi, \rpy~and \rpz~colors are very weak and ignored.
Their fitting coefficients are given in Figure\,\ref{Fig:regc}.

Figure\,\ref{Fig:delmag_radec_ini} shows the spatial variations of magnitude offsets after binning with a $20' \times 20'$ window for the \grizy~bands. We can see both large-scale and small-scale patterns 
for all the five bands. The small-scale patterns, having a typical size similar to the 
PS1 field-of-view (3.3 degree), is more clearly displayed in 
Figure\,\ref{Fig:delmag_redec_sub} for the $i$ band. The same mottling also appears in Figure\,10 of \citet{2020ApJS..251....6M}. The large-scale patterns are different between different bands. 
The $y$ band shows the strongest patterns, probably due to 
the strong and variable water absorption of the atmosphere. 
The spatial patterns suggest that the magnitude offsets are caused by calibration 
errors in the PS1 rather than Gaia.

To investigate whether the derived magnitude offsets are affected by possible systematic 
errors in reddening correction, the correlations between the magnitude offsets of different bands 
are calculated and shown in Figure\,\ref{Fig:R}. The correlation coefficients are very close to zero, 
suggesting that systematic errors in reddening correction are very small.
However, we note that the magnitude offsets of two bands can show clear correlations 
in certain sky areas. %, e.g., at (R.A., Dec.) = ()

To quantitatively estimate calibration errors of the PS1 data, 
we plot the magnitude offsets as a function 
of star numbers in one box in Figure\,\ref{Fig:conv}. The standard deviations are also estimated 
using Gaussian fitting. The values decrease as the star numbers increases first, then become flat 
when the star numbers are larger than 10. The convergence values are respectively 
4.6, 3.9, 3.5, 3.6, and 4.8 mmag 
for the \grizy~bands, confirming that the PS1 data has achieved an internal precision of $<1\%$ mag (\citealt{2012ApJ...756..158S}).

To correct the above patterns, we perform an adaptive median smoothing. 
The initial box size is $20' \times 20'$. If the star numbers within a box is less than 20, 
then the box size is doubled until it reaches to $160'$. 
The results of the five bands after smoothing are plotted in Figure\,\ref{Fig:delmag_radec_c}, and can be used to correct calibration errors in the PS1 data within the LAMOST footprint. 
The data is publicly available\footnote{\url{http://paperdata.china-vo.org/Xiao.Kai/PS1/spatial\_corr.zip}}. The corrected magnitude ${\bf M}^{\rm corr}$ can be computed 
as 
  \begin{eqnarray}
  {\bf M}^{\rm corr}={\bf M}^{\rm obs}+\Delta {\bf M}\rm (RA, Dec)+\Delta {\bf M}\rm (Mag)~,  \label{corr}
  \end{eqnarray}
where ${\bf M}^{\rm obs}$ is the observed magnitude, $\Delta {\bf M}\rm (RA, Dec)$ is the 
position-dependent magnitude offset (see Figure\,\ref{Fig:delmag_radec_c}), and $\Delta {\bf M}\rm (Mag)$ is the magnitude-dependent magnitude offset (see Table\,\ref{tab:3}).
To check the effect of correction, we over-plot the magnitude offsets after correction as a function of star numbers in Figure\,\ref{Fig:conv}. 
The standard deviations decrease to 0.4 -- 0.5 mmag for the five bands.

We also apply our corrected magnitudes of PS1 back to the linear regression process of reddening coefficients. All the fitting residuals are smaller, decreasing from 9.5 to 8.3, 
9.1 to 8.2, 7.4 to 6.8, 8.1 to 7.3, and 11.0 to 9.5 mmag for the $grizy$ bands, respectively.
These numbers are consistent with those in Figure\,\ref{Fig:conv}. For example, $\sqrt{9.5^2 - 8.3^2}$ $\sim$ 4.6.

% ==============================================================================
\section{Discussions} \label{sec:discussion}

After correcting for the magnitude-dependent errors, 
we plot variation of magnitude offsets with magnitudes and \bprp~color of the calibration samples
in Figure\,\ref{Fig:delmag_paras}.
As expected, no dependence on magnitudes is found. The dependence on \bprp~is also ignorable. 

To further verify the magnitude-dependent corrections, we perform an independent check using 
data from the re-calibrated SDSS Stripe 82 standard stars catalog (V4.2; \citealt{Huang}) and DES DR1. 
Stars within a sub-stripe of $|\rm Dec|<1.266^{\circ }$ and $23^{\rm h}00^{\rm m}<\rm {RA}<2^{\rm h}40^{\rm m}$ are used. All these stars are in the high Galactic latitude region and suffer very low extinction. 
First, we construct different color-color relations, 
e.g., $r_{\rm SDSS}-r$ versus $g_{\rm SDSS}-i_{\rm SDSS}$, and use 
them to obtain the predicted PS1 magnitudes from the SDSS/DES magnitudes and colors. 
Note that the filter differences between different surveys are corrected here via color-color relations.
Then we plot the offsets between the predicted and observed PS1 magnitudes as a function of 
PS1 magnitude. An example of the magnitude offset in $r$ band from the SDSS Stripe 82 
varying with $r$ is shown in Figure\,\ref{Fig:slr}. % Due to the different samples between the 
%Gaia, SDSS, and DES, a relative system offset are appeared. After elimination of the zero point, 
The results are summarized in Figure\,\ref{Fig:del_mag2}. Note that the results 
from the SDSS and DES are shifted slightly in the vertical direction for easy comparison. 

Figure\,\ref{Fig:del_mag2} shows that for the $g$, $r$, and $i$ bands, both the results of SDSS stripe\,82 and DES DR1 agree well with our corrections from Gaia EDR3 for magnitudes between 14 -- 17. 
For the $z$ band, our corrections are consistent with those of DES DR1 only.
%It suggests that the $z_{\rm SDSS}$ magnitude have an imperfect photometric calibration.
For the $y$ band, only result from the DES DR1 is obtained, with a much deeper slope.
The results suggest that the moderate magnitude-dependent errors in the PS1 magnitudes are real. 

To investigate the possible causes of the magnitude-dependent errors,
we select 20,000 stars from PS1 DR1 and compare their aperture-based and PSF magnitudes.
Note that all the magnitudes mentioned earlier refer to the PSF magnitudes in this work.
The results are plotted in Figure\,\ref{Fig:magap_mag}.
These results are also plotted in Figure\,\ref{Fig:del_mag2} with green dots for comparison, after shifting of the zero points. Figure\,\ref{Fig:del_mag2} suggests that our corrections from Gaia EDR3
agree with the differences between the aperture-based and PSF magnitudes in each band.
It implies that most of the magnitude-dependent errors in the PS1 PSF magnitudes probably come from 
systematic errors in the PSF magnitudes. Note that the uncertainties in the non-linearity 
corrections of the CCDs may also contribute partly. 

\citet{2020AJ....159..165P} noted that the PSF photometry using maximum-likelihood methods 
systematically overestimate the flux, with a bias scaling with the inverse signal-to-noise ratio and the number of model parameters involved in the fit. One expects a 1 per cent bias for a 10$\sigma$ 
point source and 0.01 per cent bias for a 100$\sigma$ point source. 
Given the brightness of the calibration stars used in this work, 
such effect may contribute only a small fraction to the systematic errors in the PS1 PSF magnitudes.

\section{Conclusions} \label{sec:conclusion}
In this paper, using the SCR method with the photometric data from the corrected Gaia EDR3 and the spectroscopic data from LAMOST DR7, we have performed an 
independent validation and recalibration of the PS1 photometry. 
Using typically a total of 1.5 million LAMOST-PS1-Gaia FGK dwarf stars as standards per band, 
we show that the PS1 photometric calibration precisions are respectively 4.6, 3.9, 3.5, 3.6, and 4.8 mmag in
the $grizy$ bands when averaged over $20'$ regions.
However, significant large- and small-scale spatial variation of magnitude offsets, 
up to over 1 per cent and caused by calibration errors in the PS1, are found for all the $grizy$ filters.
The calibration errors in different filters are un-correlated in most sky areas.
The $y$ band shows the strongest patterns, probably due to 
the strong and variable water absorption of the atmosphere. 

We also detect the moderate magnitude-dependent errors in the PS1 photometry, 
i.e., 0.005, 0.005, 0.005, 0.004, 0.003 mag per magnitude in the 14 -- 17 magnitude range 
for the $grizy$ filters, respectively. Such errors are further confirmed by comparing the PS1 magnitudes with 
those predicted from the re-calibrated SDSS Stripe 82 standard stars catalog (V4.2) and DES DR1. Such errors are likely caused by systematic uncertainties in the PSF magnitudes.
%non-linearity correction of the PS1 CCDs.

It implies that most of the magnitude-dependent errors in the PS1 PSF magnitudes probably come from 
systematic errors in the PSF magnitudes. Note that the uncertainties in the non-linearity 
corrections of the CCDs may also contribute partly. 

We provide two-dimensional maps to correct for position-dependent magnitude offsets in the LAMOST footprint at resolutions from $20'$ to $160'$. The maps, together with the magnitude-dependent corrections, 
are useful in the usage of the PS1 photometry for high-precision investigations (e.g., 
\citealt{2021ApJ...922..211N, 2021arXiv211111725X}) and 
as the reference to calibrate other surveys.

The results demonstrate the power of the SCR method in improving calibration precision 
of wide-field surveys when combined with Gaia photometry. We suggest that the SCR method should be
incorporated into the calibration process of future releases of the PS1 data.

\begin{acknowledgments}

%We acknowledge the anonymous referee for his/her valuable comments that improve the quality of this paper.
We thank Prof. Xiaowei Liu for valuable discussions and a careful reading of the manuscript. 
This work is supported by the National Natural Science Foundation of China through the project NSFC 12173007 and 11603002,
the National Key Basic R\&D Program of China via 2019YFA0405503 and Beijing Normal University grant No. 310232102. 
We acknowledge the science research grants from the China Manned Space Project with NO. CMS-CSST-2021-A08 and CMS-CSST-2021-A09.

This work has made use of data from the European Space Agency (ESA) mission Gaia (\url{https://www.cosmos.esa.int/gaia}), processed by the Gaia Data Processing and Analysis Consortium (DPAC, \url{https:// www.cosmos.esa.int/web/gaia/dpac/ consortium}). Funding for the DPAC has been provided by national institutions, in particular the institutions participating in the Gaia Multilateral Agreement. Guoshoujing Telescope (the Large Sky Area Multi-Object Fiber Spectroscopic Telescope LAMOST) is a National Major Scientific Project built by the Chinese Academy of Sciences. Funding for the project has been provided by the National Development and Reform Commission. LAMOST is operated and managed by the National Astronomical Observatories, Chinese Academy of Sciences.

The Pan-STARRS1 Surveys (PS1) and the PS1 public science archive have been made possible through contributions by the Institute for Astronomy, the University of Hawaii, the Pan-STARRS Project Office, the Max-Planck Society and its participating institutes, the Max Planck Institute for Astronomy, Heidelberg and the Max Planck Institute for Extraterrestrial Physics, Garching, The Johns Hopkins University, Durham University, the University of Edinburgh, the Queen's University Belfast, the Harvard-Smithsonian Center for Astrophysics, the Las Cumbres Observatory Global Telescope Network Incorporated, the National Central University of Taiwan, the Space Telescope Science Institute, the National Aeronautics and Space Administration under Grant No. NNX08AR22G issued through the Planetary Science Division of the NASA Science Mission Directorate, the National Science Foundation Grant No. AST–1238877, the University of Maryland, Eotvos Lorand University (ELTE), the Los Alamos National Laboratory, and the Gordon and Betty Moore Foundation.

\end{acknowledgments}

\end{document}